\documentclass[12pt]{elsarticle}
\usepackage{amsmath,amsfonts}
\usepackage{algorithmic,bm}
\usepackage{hyperref}
\usepackage{mathtools}
\usepackage{algorithm}
\usepackage{array}
\usepackage[caption=false,font=normalsize,labelfont=sf,textfont=sf]{subfig}
\usepackage{multirow}
\usepackage{textcomp}
\usepackage{stfloats}
\usepackage{url}
\usepackage{verbatim}
\usepackage{graphicx}
\usepackage{xcolor}
\usepackage{accents}
\usepackage{bbm}
\usepackage{optidef}
\usepackage{rotating}
\usepackage{siunitx}
\usepackage{arydshln}
\usepackage{enumitem}
\usepackage{epstopdf}
\usepackage{setspace}

\usepackage{ifthen}
\newboolean{commentOut}
\setboolean{commentOut}{true}

\ExplSyntaxOn
\NewDocumentCommand{\longdash}{ O{2} }
 {
  --\prg_replicate:nn { #1 - 1 } { \negthinspace -- }
 }
\ExplSyntaxOff

\newcommand{\NA}{---}

\newcommand{\llevl}{\mathbbm{L}}
\newcommand{\ulevl}{\mathbbm{U}}

\allowdisplaybreaks

\journal{Electric Power Systems Research}

\makeatletter
\def\ps@pprintTitle{%
     \let\@oddhead\@empty
     \let\@evenhead\@empty
     \def\@oddfoot{\reset@font\hfil\thepage\hfil}
     \let\@evenfoot\@oddfoot
}
\makeatother

\begin{document}
\doublespacing
\begin{frontmatter}
\title{A Data-Driven Sensor Placement Approach for Detecting Voltage Violations in Distribution Systems}

\author[1]{Paprapee Buason\corref{cor1}}
\ead{pbuason6@gatech.edu}
\cortext[cor1]{Corresponding author}

\author[2]{Sidhant Misra}
\ead{sidhant@lanl.gov}

\author[1]{Samuel Talkington}
\ead{talkington@gatech.edu}

\author[1]{Daniel K. Molzahn}
\ead{molzahn@gatech.edu}

\address[1]{Georgia Institute of Technology}
\address[2]{Los Alamos National Laboratory}

\begin{abstract}
Stochastic fluctuations in power injections from distributed energy resources (DERs) combined with load variability can cause constraint violations (e.g., exceeded voltage limits) in electric distribution systems. To monitor grid operations, sensors are placed to measure important quantities such as the voltage magnitudes. In this paper, we consider a sensor placement problem which seeks to identify locations for installing sensors that can capture all possible violations of voltage magnitude limits. We formulate a bilevel optimization problem that minimizes the number of sensors and avoids false sensor alarms in the upper level while ensuring detection of any voltage violations in the lower level. This problem is challenging due to the nonlinearity of the power flow equations and the presence of binary variables. Accordingly, we employ recently developed conservative linear approximations of the power flow equations that overestimate or underestimate the voltage magnitudes. By replacing the nonlinear power flow equations with conservative linear approximations, we can ensure that the resulting sensor locations and thresholds are sufficient to identify any constraint violations. Additionally, we apply various problem reformulations to significantly improve computational tractability while simultaneously ensuring an appropriate placement of sensors. Lastly, we improve the quality of the results via an approximate gradient descent method that adjusts the sensor thresholds. We demonstrate the effectiveness of our proposed method for several test cases, including a system with multiple switching configurations.
\end{abstract}

\begin{keyword}
Keywords\sep Sensor placement\sep voltage violations\sep bilevel optimization\sep approximate gradient descent.
\end{keyword}

\end{frontmatter}

\section{Introduction}
Distributed energy resources (DERs) are being rapidly deployed in distribution systems. Fluctuations in DER power outputs and varying load demands can potentially cause violations of voltage limits, i.e., voltages outside the bounds imposed in the ANSI C84.1 standard. These violations can cause equipment malfunctions, failures of electrical components, and, in severe situations, power outages.

To mitigate the impacts of violations, distribution system operators (DSOs) must identify when power injection fluctuations lead to voltages exceeding their limits. To do so, sensors are placed within the distribution system to measure and communicate the voltage magnitudes at their locations. Due to the cost of sensor hardware and communication infrastructure and the structure of distribution systems, sensors are not placed at all buses. The question arises whether a voltage violation at a location without a sensor can nevertheless be detected.

\subsection{Related work}

Various studies have proposed sensor siting methods to capture constraint violations and outages in power systems. For instance,~\cite{SAMUDRALA2019} and~\cite{SAMUDRALA2020} focus on a cost minimization problem that aims to capture all node (e.g., voltage magnitude) and line (e.g., power flow) outages. However, these references assume that a power source/generation is only located at the root node, which is not always the case, especially in the distribution systems where DERs can be located further down a feeder. Additional research efforts such as 
\cite{BALDWIN1993, GOU2008, ALBUQUERQUE2013, THOMAS2014} 
seek to locate the minimum number of sensors to achieve full observability for the system. Alternatively, instead of considering full observability for the entire system, \cite{Aminifar2011} considers satisfying observability requirements given a probability of observability at each bus. This is related to adjacent research efforts, such as \cite{MEHRJERDI2013, NOUR2020, Dehkordi2020, Soliman2023}, which focus on model-based voltage control schemes that prevent voltage violations. We refer the reader to \cite{murray2021voltage} for a survey.
In parallel, there is also research on siting phasor measurement units (PMUs) and micro phasor measurement units ($\mu$-PMUs), which are utilized as sensors in power grids~\cite{Mohanta2016, Jamei2018, Dusabimana2020}. This research complements the discussion on sensor placement strategies by exploring alternative sensor technologies and their deployment strategies within power systems. Existing smart meters installed at customer locations measure power consumption over a long period of time (e.g., a day to a month). Reference~\cite{McKenna_2022} discusses the potential benefits of using smart meters to report voltages and currents at higher resolutions. However, handling large amounts of data remains challenging, necessitating methods to selectively site high-resolution sensors.

Growing numbers of sensing devices in electric distribution networks have fueled a broad range of applications. Exemplifying use cases of the measurements from these sensors are in control \cite{murray2021voltage,wang2018distributed,gupta2022model}, state estimation \cite{Baran1994, Primadianto2017,melo_neutral--earth_2023}, topology identification  \cite{stanojev_tractable_2023,deka_learning_2024}, and non-intrusive load monitoring (NILM) \cite{schirmer_nilm_review_2023}. Our work contributes to this rich literature. Some measurement-based voltage control approaches, such as \cite{wang2018distributed,gupta2022model} utilize approximate representations of the power flow equations recovered from data; like our work, this approach is an abstract way to model an unknown distribution network. In contrast, network identification works such as \cite{stanojev_tractable_2023,vanin2023combined} seek to recover the network topology itself, in tandem with admittance \cite{stanojev_tractable_2023} or impedance \cite{vanin2023combined} parameters. We refer the reader to \cite{deka_learning_2024} for a comprehensive review of this topic.

\subsection{Overview of approach}
In this paper, we consider a \emph{sensor placement} problem which seeks to locate the minimum number of sensors and determine corresponding sensor alarm thresholds in order to reliably identify all possible violations of voltage magnitude limits in a distribution system. We formulate this sensor placement problem as a bilevel optimization with an upper level that minimizes the number of sensors and chooses sensor alarm thresholds and a lower level that computes the most extreme voltage magnitudes within given ranges of power injection variability. This problem additionally aims to reduce the number of false positive alarms, i.e., violations of the sensors' alarm thresholds that do not correspond to an actual voltage limit violation. 

In contrast to previous work, this problem \emph{does not} attempt to ensure full observability of the distribution system. Rather, we seek to locate (a potentially smaller number of) sensors that can nevertheless identify all voltage limit violations for any power injections within a specified range of power injection variability. With a small number of sensors, the proposed formulation also provides a simple means to design corrective actions if voltage violations are encountered in real-time operations. By restoring voltages at these few critical locations to within their alarm thresholds, the system operator can guarantee feasibility of the voltage limits for the full system. This guarantee is obtained by our sensor placement method purely by analyzing the geometric properties of the feasible set. We do not consider the design details of the feedback control protocol and thus dynamic properties of the sensors such as latency are not relevant in our approach. 

Due to the nonlinear nature of the AC power flow equations, computing a globally optimal solution is challenging. We utilize \textit{conservative linear approximations} of the power flow equations to convert the lower-level problem to a linear program~\cite{BUASON2022}. This bilevel problem can be reformulated to a single-level problem using the Karush-Kuhn-Tucker (KKT) conditions with binary variables via a big-M formulation~\cite{WEN1991, dempe2012}.
In this paper, we consider a duality-based approach, which has substantial computational advantages over traditional KKT-based approaches to solving the bilevel problem. \textcolor{black}{The conservative linear approximations can incorporate the behavior of more complex components such as tap-changing transformers, smart inverters, etc., as long has we have access to a power flow solver for the system. By using these linear approximations as the first step, we are able to treat the power flow solver as a black-box. Consequently, all complexities of component behavior and power flow physics are absorbed by the power flow solver and the complexity of the resulting sensor placement formulation remains unaffected.}

Note that conservativeness from the conservative linear approximations may increase the number of false positive alarms. We therefore propose an approximate gradient descent method as a post-processing step to further improve the quality of the results. This method iteratively adjusts the sensor thresholds while ensuring that all violations are still detected.

\subsection{Contributions and article outline}

In summary, our main contributions are:
\begin{enumerate}[label=(\textit{\roman*})]
    \item A bilevel optimization formulation for a sensor placement problem that minimizes the number of sensors needed to capture all possible violations of voltage limits while minimizing the number of false positive alarms. 
    \item Reformulations that substantially improve the computational tractability of this bilevel problem.
    \item An approximate gradient descent method to improve solution quality.
    \item Numerical demonstration of our proposed problem formulations for a variety of test cases, including networks with multiple switching configurations. 
\end{enumerate}

This paper is organized as follows. Section~\ref{sec:problem_formulation} formulates the sensor placement problem using bilevel optimization. Section~\ref{sec:problem_reformulation} proposes different techniques to reformulate the optimization problem. Section~\ref{sec:simulation} provides our numerical tests. Section~\ref{sec:conclusion} concludes the paper and discusses future work.

\section{Sensor Placement Problem} \label{sec:problem_formulation}
This section describes the sensor placement problem by introducing notation, presenting the bilevel programming formulation that is the focus of this paper, and detailing the objective function that simultaneously minimizes the number of sensors and reduces the number of false positive alarms. 

\subsection{Notation}
Consider an $n$-bus power system. The sets of buses and lines are denoted as $\mathcal{N} = \{1,\ldots, n\}$ and~$\mathcal{L}$, respectively. One bus in the system is specified as the slack bus where the voltage is $1 \angle 0$\textdegree~per unit. For the sake of simplicity, the remaining buses are modeled as PQ buses with given values for their active ($P$) and reactive ($Q$) power injections. Extensions to consider PV buses, which have given values for the active power ($P$) and the voltage magnitude ($V$), are straightforward. (The controlled voltage magnitudes at PV buses imply that voltage violations cannot occur at these buses so long as the voltage magnitude setpoints are within the voltage limits.)  The set of all nonslack buses is denoted as $\mathcal{N}_{PQ}$. The set of neighboring buses to bus $i$ is defined as $\mathcal{N}_i := \left \{k \, | \, (i,k) \in \mathcal L \right \}$.  Subscript $(\cdot)_{i}$ denotes a quantity at bus $i$, and subscript $(\cdot)_{ik}$ denotes a quantity associated with the line from bus $i$ to bus $k$, unless otherwise stated. Conductance (susceptance) is denoted as $G$ $(B)$ as the real (imaginary) part of the admittance.

To illustrate the main concepts in this paper, we consider a balanced single-phase equivalent network representation rather than introducing the additional notation and complexity needed to model an unbalanced three-phase network. Our work does not require assumptions regarding a radial network structure, and we are able to handle multiple network configurations.
Extensions to consider other limits such as restrictions on line flows,  
budget uncertainty sets, and unbalanced three-phase network models impose limited additional complexity.

\subsection{Bilevel optimization formulation}
\par The main goal of this problem is to find sensor location(s) such that sensor(s) can capture all possible voltage violations. This paper assumes that the voltages read by the sensors are accurate and noise-free measurements. We formulate this problem as a bilevel optimization with the following upper-level and lower-level problems.
\begin{itemize}
    \item \textbf{Upper level:} Determine sensor locations and alarm thresholds such that when the voltages at the sensors are within the chosen thresholds, the voltages at all other buses are within pre-specified safety limits.
    \item \textbf{Lower level:} Find the extreme achievable voltages at all buses given the sensor locations, sensor alarm thresholds, and the specified range of power injection variability.
\end{itemize}

The sensor locations and alarm thresholds output from the upper-level problem are input to the lower-level problem, and the extreme achievable voltage magnitudes output from the lower-level problem are used to evaluate the bounds in the upper-level problem. 
We first introduce notation for various quantities associated with the voltage at bus $i$:\vspace{-1em}
\begin{align*} 
&\undertilde{V}_i\; (\widetilde{V}_i) &&: \text{Lower (Upper) sensor alarm threshold.}\\[-0.25em]
&\underline{V}_i\; (\overline{V}_i) &&: \text{Lowest (Highest) achievable voltage}\\[-0.25em]
& &&\hphantom{:} \hspace{0.2cm} \text{obtained from the lower-level problem.}\\[-0.25em]
&\undertilde{U}_i\; (\widetilde{U}_i) &&: \text{Translation of lower (upper) sensor}\\[-0.25em]
& &&\hphantom{:} \hspace{0.2cm} \text{threshold via a big-M formulation; see~\eqref{sens_limit_upper_full}.}\\
&V_i^{\text{min}}\; (V_i^{\text{max}}) &&: \text{Specified lower (upper) voltage limit.}
\end{align*}
We formulate the following bilevel optimization formulation:
\begin{subequations} \label{full_problem}
\begin{align} 
    \min_{\mathbf{s},\widetilde{\mathbf{V}}, \undertilde{\mathbf{V}}} \enspace & c(\mathbf{s},\widetilde{\mathbf{V}}, \undertilde{\mathbf{V}}) \label{upper_level_objective_full} \\
    \text{s.t.} \enspace & (\forall i \in \mathcal{N}_{PQ}) \nonumber \\
    & \underline{V}_i \geq V_i^{\text{min}}, \; \overline{V}_i \leq V_i^{\text{max}}, \label{threshold_full} \\
    & \undertilde{U}_{i} = \undertilde{V}_i s_i, \ \widetilde{U}_{i} = \widetilde{V}_i s_i + M (1-s_i), \label{sens_limit_upper_full} \\
    & \underline{V}_i = \llevl_i(\mathbf{s},\widetilde{\mathbf{U}},\undertilde{\mathbf{U}}), \; \overline{V}_i = \ulevl_i(\mathbf{s},\widetilde{\mathbf{U}},\undertilde{\mathbf{U}}), \label{upper_lower_notation} 
    \end{align}
\end{subequations}
where $c$ is the cost function associated with the placement of sensors including costs for the hardware, installation, communication network, etc. and $\mathbf{s}$ is a vector of sensor locations modeled as binary variables (1 if a sensor is placed, 0 otherwise). All bold quantities are vectors. 
The quantities $\llevl_i(\mathbf{s},\widetilde{\mathbf{U}},\undertilde{\mathbf{U}})$ and $\ulevl_i(\mathbf{s},\widetilde{\mathbf{U}},\undertilde{\mathbf{U}})$ are the solutions to the lower-level problems which, for each $i \in \mathcal{N}_{PQ}$, are given by
\begin{subequations} \label{lower_level}
\begin{align} 
        \llevl_i&(\mathbf{s},\widetilde{\mathbf{U}},\undertilde{\mathbf{U}}) = \ \min_{V_i} \enspace  V_i \quad (\ulevl_i(\mathbf{s},\widetilde{\mathbf{U}},\undertilde{\mathbf{U}}) = \ \max_{V_i} \enspace V_i)\label{lower_level_objective_full}  \\
        & \textrm{s.t.} \enspace (\forall j \in \mathcal{N}_{PQ}) \nonumber \\
        & P_j = V_j\sum_{k \in \mathcal{N}_j}V_k (G_{jk}\cos (\theta_{jk}) + B_{jk}\sin(\theta_{jk})), \label{PF_P} \\
        & Q_j = V_j\sum_{k \in \mathcal{N}_j}V_k (G_{jk}\sin (\theta_{jk}) - B_{jk}\cos(\theta_{jk})), \label{PF_Q} \\
        & \undertilde{U}_j \leq V_j \leq \widetilde{U}_j, \label{sens_limit_full} \\
        & P_j^{\text{min}} \leq P_j \leq P_j^{\text{max}}, \label{load_limit_P_full} \\
        & Q_j^{\text{min}} \leq Q_j \leq Q_j^{\text{max}}, \label{load_limit_Q_full} 
    \end{align}
\end{subequations}
where $P_j$  and $Q_j$ denote the active and reactive power injections at bus $j$ within a particular lower-level problem, $\theta_{jk}:=\theta_j - \theta_k$ denotes the voltage angle difference between buses~$j$ and $k$, and superscripts max and min denote upper and lower limits, respectively, on the corresponding quantity. The quantities $\llevl_i$ and $\ulevl_i$ are functions of $\mathbf{s}$, $\widetilde{\mathbf{U}}$, and $\undertilde{\mathbf{U}}$ as shown in~\eqref{upper_lower_notation}, but these dependencies are hereafter omitted for the sake of notational brevity. For the upper-level problem, the objective function in \eqref{upper_level_objective_full} minimizes a cost function $c(\mathbf{s},\widetilde{\mathbf{V}}, \undertilde{\mathbf{V}})$ associated with the sensor locations $\mathbf{s}$ and alarm thresholds $\widetilde{\mathbf{V}}$, $\undertilde{\mathbf{V}}$ while ensuring that the extreme achievable voltage magnitudes calculated in the lower-level problem, $\underline{V}_i,\overline{V}_i$ are within safety limits as shown in \eqref{threshold_full}. The cost function $c(\mathbf{s},\widetilde{\mathbf{V}}, \undertilde{\mathbf{V}})$ will be detailed in the following subsection. In the lower-level problem, the objective function \eqref{lower_level_objective_full} computes the maximum or minimum voltage magnitude for each PQ bus $i\in\mathcal{N}_{PQ}$. For each lower-level problem, constraints \eqref{PF_P}--\eqref{PF_Q} are the power flow equations at each bus $j$, \textcolor{black}{constraint \eqref{sens_limit_full} forces the voltage magnitudes to be within voltage alarm thresholds if a sensor is placed at the corresponding bus}, and constraints \eqref{load_limit_P_full}--\eqref{load_limit_Q_full} model the range of variability in the net power injections. We typically set $\theta_1$ = 0 as the angle reference.

\subsection{Cost function} \label{sec:cost_function}
Overly restrictive sensor thresholds can potentially trigger an alarm even when there are no voltage violations actually occurring in the system, thus resulting in a \textit{false positive}. To reduce both the number of sensors and the number of false positive alarms due to unnecessarily restrictive alarm thresholds, our cost function, $c(\mathbf{s},\undertilde{\mathbf{V}},\widetilde{\mathbf{V}})$, is:
\begin{align}
    c(\mathbf{s},\undertilde{\mathbf{V}},\widetilde{\mathbf{V}}) = \sum_{i \in \mathcal{N}} c_i(s_i,\undertilde{V}_i,\widetilde{V}_i)
\end{align}
where
\begin{equation} \label{cost_function}
    c_i(s_i,\undertilde{V}_i,\widetilde{V}_i) = 
    \begin{cases}
    (\undertilde{V}_i - {V}_i^{\text{min}}) + ({V}_i^{\text{max}} - \widetilde{V}_i) + \delta; & s_i = 1, \\
    0; & s_i = 0,
    \end{cases}
\end{equation}
where $\delta$ is a specified cost of placing a sensor. When $s_i = 1$, the objective $c(\mathbf{s},\undertilde{\mathbf{V}},\widetilde{\mathbf{V}})$ in \eqref{cost_function} seeks to reduce the restrictiveness of the sensor alarm thresholds to have fewer false positives. Changing the value of $\delta$ in \eqref{cost_function} models the tradeoff between placing an additional sensor and making the sensor range more restrictive. This is a crucial part of our formulation since our main goal is to identify a small number of critical locations that carry sufficient information about the feasibility of the entire network. Beyond the clear financial benefit of having to place fewer sensors, this also provides a simple and practical mechanism for deploying corrective actions in real-time. Indeed, when the system operator encounters a voltage violation, a reactive power compensation protocol that brings the voltages at these few critical locations to within the alarm thresholds will guarantee feasibility of the voltage limits for the entire network.

\section{Reformulations of the Sensor Location Problem} \label{sec:problem_reformulation}
The bilevel problem~\eqref{full_problem} is computationally challenging due to the non-convexity in the lower-level problem induced by the AC power flow equations in~\eqref{PF_P}--\eqref{PF_Q} and the presence of two levels. 
In this section, we provide methods for obtaining a tractable version of the bilevel sensor placement problem. We first use the \emph{conservative linear approximations} of the power flow equations to convert the lower-level problem to a more tractable linear programming formulation that nevertheless retains characteristics from the nonlinear AC power flow equations. This bilevel problem can be reformulated to a single-level problem using the Karush-Kuhn-Tucker (KKT) conditions with binary variables via a big-M formulation~\cite{WEN1991, dempe2012}. However, as we will show numerically in Section~\ref{sec:simulation}, traditional methods for reformulating the bilevel problem into a single-level problem suitable for standard solvers using the KKT conditions turn out to yield computationally burdensome problems. (The full problem setup using the KKT conditions is shown in~\ref{sec:kkt}.) We then use various additional reformulation techniques that yield significantly more tractable problems than standard KKT-based reformulations. These reformulations first yield a (single-level) mixed-integer bilinear programming formulation that can be solved using commercial mixed-integer programming solvers like Gurobi. We further discretize the sensor alarm thresholds and transform the bilinear terms to a mixed-integer linear program (MILP).

\subsection{Conservative linear power flow approximations} \label{sec:cla}
To address challenges associated with power flow nonlinearities, we employ a linear approximation of the power flow equations that is adaptive (i.e., tailored to a specific system and a range of load variability) and conservative (i.e., intend to over- or under-estimate a quantity of interest to avoid constraint violations). These linear approximations are called \textit{conservative linear approximations} (CLAs) and were first proposed in~\cite{BUASON2022}. As a sample-based approach, the CLAs are computed using the solution to a constrained regression problem across all samples within the range of power injection variability. They linearly relate the voltage magnitudes at a particular bus to the power injections at all PQ buses. These linear approximations can also effectively incorporate the characteristics of more complex components (e.g., tap-changing transformers, smart inverters, etc.), only requiring the ability to apply a power flow solver to the system. Additionally, in the context of long-term planning, the CLAs can be readily computed with knowledge of expected DER locations and their potential power injection ranges. The accuracy and conservativeness of our proposed method is based on the information of the location of DERs and their power injections variability. As inputs, our method uses the net load profiles including the size of PVs when computing the CLAs. In practice, this data can be obtained by leveraging the extensive existing research on load modeling and monitoring to identify the locations and capabilities of behind-the-meter devices (refer to, e.g.,~\cite{Grijalva2021, Schirmer2023}).

An example of an \emph{overestimating} CLA of the voltage magnitude at bus $i$ is the linear expression
\begin{equation}
a_{i,0} + \bm{a}_{i,1}^T\begin{pmatrix}
  \mathbf{P} \\
  \mathbf{Q} \nonumber
\end{pmatrix}
\end{equation}
such that the following relationship is satisfied for some specified range of power injections $\mathbf{P}$ and $\mathbf{Q}$:

\begin{equation} 
V_i - \left(a_{i,0} + \bm{a}_{i,1}^T\begin{pmatrix}
  \mathbf{P} \\  \mathbf{Q}
\end{pmatrix}\right) \leq 0,
\label{eq:cla}
\end{equation}
where $a_0$ and $\bm{a}_1$ are the coefficients of the CLA computed by solving a constrained regression problem (refer to~\cite{BUASON2022}) and  superscript $T$ denotes the transpose. We replace the AC power flow equations in \eqref{PF_P}--\eqref{PF_Q} with a CLA as in \eqref{eq:cla} for all $i \in \mathcal{N}_{PQ}$.
The quantities $\llevl_i$ and $\ulevl_i$ in~\eqref{lower_level_objective_full} become:
\begin{align} 
        \llevl_i = \min_{\mathbf{P},\mathbf{Q}} \enspace & \underline{a}_{i,0} + \underline{\bm{a}}_{i,1}^{T} 
        \begin{pmatrix}
        \mathbf{P} \\ \mathbf{Q} 
        \end{pmatrix}^i  \;
        \bigg(\ulevl_i = \max_{\mathbf{P},\mathbf{Q}} \enspace \overline{a}_{i,0} + \overline{\bm{a}}_{i,1}^{T} 
        \begin{pmatrix}
        \mathbf{P} \\ \mathbf{Q} 
        \end{pmatrix}^i\bigg) \label{lower_level_objective_cla}
\end{align}
where superscripts $i$ denote quantities associated with the $i^{\text{th}}$ lower-level problem. 
Using conservative linear approximations yields a linear programming formulation for the lower-level problem rather than the non-convex lower-level problem in~\eqref{lower_level}. By assuming that the conservative linear approximations do indeed reliably over- or under- estimate the voltage magnitudes, it is sufficient to ensure satisfaction of~\eqref{threshold_full}. As a result, solving the reformulationwill compute sensor locations and thresholds where alarms will always be raised if there are indeed violations of the voltage limits.

\subsection{Duality of the lower-level problem}\label{subsec:duality}

One can reformulate a bilevel problem into a single-level problem by \emph{dualizing} the lower-level problem. This technique can only be usefully applied to problems with specific structure where the optimal objective value of the lower-level problem is constrained in the upper-level problem in the appropriate sense ($\max \leq \cdot$ or $\min \geq \cdot$). In this special case, we can significantly improve tractability compared to the KKT formulation.

Let $\undertilde{\mathbf{y}}^i$ be the vector of all dual variables associated with the lower-level problem $\llevl_i$ and $\widetilde{\mathbf{y}}^i$ be the vector of all dual variables associated with lower-level problem $\ulevl_i$. 
Let $I$ be the identity matrix of appropriate dimension. By dualizing the lower-level problem \eqref{lower_level_objective_cla} with conservative linear power flow approximations as constraints, we obtain the following: 

\begin{subequations} 
\noindent\centering
\begin{minipage}{0.48\textwidth} \label{lower_level_objective_dual_l}
\begin{align}
\llevl_i = \max_{\undertilde{\mathbf{y}}^i} \enspace & \mathbf{b}^T \undertilde{\mathbf{y}}^i + \underline{a}_{i,0} \label{dual_problem_l} \\
\textrm{s.t.} \enspace & A\undertilde{\mathbf{y}}^i = \underline{\bm{a}}_{i,1}, \label{power_injection_dual_l} \\
& \undertilde{\mathbf{y}}^i \geq \mathbf{0}, \label{dual_var_l} 
\end{align}
\end{minipage}%
\end{subequations}%
\hfill
\begin{subequations}%
\begin{minipage}{0.48\textwidth}\label{lower_level_objective_dual_u}%
\begin{align}%
\ulevl_i = \min_{\widetilde{\mathbf{y}}^i\vphantom{\undertilde{\mathbf{y}}^i}} \enspace & \mathbf{b}^T \widetilde{\mathbf{y}}^i + \overline{a}_{i,0} \label{dual_problem_u} \\
\textrm{s.t.} \enspace & A\widetilde{\mathbf{y}}^i = \overline{\bm{a}}_{i,1}, \label{power_injection_dual_u} \\
& \widetilde{\mathbf{y}}^i \leq \mathbf{0}, \label{dual_var_u}
\end{align}
\end{minipage} \bigskip
\end{subequations}
where 
\begin{align*}
    A &= 
    \left[
    -I, I, \overline{\bm{a}}_{1,1}, \hdots, \overline{\bm{a}}_{n,1},-\underline{\bm{a}}_{1,1}, \hdots, -\underline{\bm{a}}_{n,1}
    \right], \\
    \mathbf{b} &= [ 
    (-\mathbf{P}^{\text{max}})^T, (-\mathbf{Q}^{\text{max}})^T, (\mathbf{P}^{\text{min}})^T, (\mathbf{Q}^{\text{min}})^T,  
    \undertilde{U}_{1} - \overline{a}_{1,0}, \hdots, \undertilde{U}_{n} - \overline{a}_{n,0}, \\ & \qquad - \widetilde{U}_{1} + \underline{a}_{1,0}, \hdots, - \widetilde{U}_{n} + \underline{a}_{n,0} 
    ]^T.
\end{align*}

Due to strong duality of the linear lower-level problem, the dual~\eqref{dual_problem_l} (and \eqref{dual_problem_u}) has the same objective value as~\eqref{lower_level_objective_cla} and does not directly provide any advantages. However, the problem has a specific structure where objectives from each lower-level problem~\eqref{dual_problem_l} and \eqref{dual_problem_u} only appear in a single inequality constraint \eqref{threshold_full}. Hence, we \emph{only need to show that there exists some choice of duals $\undertilde{\mathbf{y}}^i$ and $\widetilde{\mathbf{y}}^i$  for which \eqref{threshold_full} is feasible}. This allows us to obtain a single-level formulation by defining the lower-level coupling quantities via the following set of constraints:

\vspace{-1em}
\begin{subequations}   
\noindent\centering
\begin{minipage}{0.48\textwidth}  \label{billinear_lower_l}
\begin{align}
    \llevl_i = \ &\mathbf{b}^T \undertilde{\mathbf{y}}^i + \underline{a}_{i,0},  \label{lower_level_objective_dual_single_l} \\
    &A\, \undertilde{\mathbf{y}}^i = \underline{\bm{a}}_{i,1}, \label{power_injection_dual_single_l} \\
    &\undertilde{\mathbf{y}}^i \geq \mathbf{0}. \label{dual_var_single_l}
\end{align}
\end{minipage}
\end{subequations}
\hfill
\begin{subequations}
\begin{minipage}{0.48\textwidth} \label{billinear_lower_u}
\begin{align}
    \ulevl_i = \ &\mathbf{b}^T \widetilde{\mathbf{y}}^i + \overline{a}_{i,0},  \label{lower_level_objective_dual_single_u} \\
    &A\, \widetilde{\mathbf{y}}^i = \overline{\bm{a}}_{i,1}, \label{power_injection_dual_single_u} \\
    &\widetilde{\mathbf{y}}^i \leq \mathbf{0}. \label{dual_var_single_u}
\end{align}
\end{minipage} \bigskip
\end{subequations}

We refer to the formulation using~\eqref{billinear_lower_l} and \eqref{billinear_lower_u} as the ``bilinear formulation'' due to the bilinear product of the sensor threshold variables ($\undertilde{\mathbf{U}}$ and $\widetilde{\mathbf{U}}$) and the dual variables $\undertilde{\mathbf{y}}^i$ and $\widetilde{\mathbf{y}}^i$ in~\eqref{lower_level_objective_dual_single_l} and \eqref{lower_level_objective_dual_single_u}. Using~\eqref{billinear_lower_l} and~\eqref{billinear_lower_u} leads to a single-level optimization problem. However, the latter has the major advantage that no additional binary variables are required (beyond those associated with the sensor locations in the upper-level problem) since there are no analogous equations to the complementarity condition as in the KKT reformulation. Our bilinear formulation can be further converted into an MILP by discretizing the continuous-valued sensor thresholds. The details for this MILP reformulation and the removal of unnecessary binary variables from discretizing the sensor thresholds (referred to as binary variable removal (BVR)) are described in~\ref{subsec:milp_reformulation}. Further details about the comparison of each problem formulation provided in this paper, including the use of the KKT conditions, are in~\ref{sec:comparison}.

\subsection{Approximate gradient descent} \label{sec:agd}
Solving any of the reformulated bilevel optimization problems may lead to \emph{false positives}.
This is both due to the limited number of sensors and the conservative nature of the linear power flow approximations used in the lower-level problem. 
To reduce the number of false positives, we propose a post-processing step that iteratively adjusts the sensor thresholds that result from the reformulated bilevel optimization problems. We refer to this post-processing step as the Approximate Gradient Descent (AGD) method. 

Let superscript $k$ denote the $k^{\text{th}}$ iteration of the AGD method. Let $\epsilon_{AGD}$ be a step size for adjusting the sensor thresholds and $\mathbf{f}^k$ be the vector of the number of false positives from the sampled power injections.
Using the sampled power injections, this method computes an ``approximate gradient'' indicating how small changes to the sensor alarm thresholds affect the number of false positives. The approximate gradient at iteration $k$ is given by $\mathbf{g}^k$. We denote the set of buses with sensors as $\mathcal{N}_{s}$. Subscripts give the bus number.

Let $\Delta f^k_i$ represent the change in the number of false positives among the sampled power injections using the sensor thresholds in the $k^{\text{th}}$ iteration when the sensor alarm threshold~$i$
is changed by $\epsilon_{AGD}$ (leaving all other sensor thresholds unchanged). We then compute an approximate gradient $\mathbf{g}^k$ by comparing the values of $\Delta f^k_i$ across different buses $i$:
\begin{align}
\mathbf{g}^k &= \cfrac{\Delta \mathbf{f}^k}{\sqrt{\sum_{i \in \mathcal{N}_s} (\Delta f^k_i)^2}}.
\end{align}

In each iteration, we update the sensor thresholds as follows:
\begin{align} \label{eq:AGD_update}
\undertilde{\mathbf{V}}^{k + 1} =\undertilde{\mathbf{V}}^{k} + \epsilon_{AGD} \cdot \mathbf{g}^k \quad (\widetilde{\mathbf{V}}^{k + 1} =\widetilde{\mathbf{V}}^{k} + \epsilon_{AGD} \cdot \mathbf{g}^k).
\end{align}

The AGD method stops when taking an additional step would result in the appearance of false \emph{negatives}, i.e., undetected violations of voltage limits. 
\section{Numerical Tests} \label{sec:simulation}
This section describes numerical experiments on a number of test cases to analyze the sensor locations and thresholds, demonstrate the advantages of our problem reformulations and the post-processing step, and compare results and computational efficiency from different problem formulations. 
\par The test cases we use in these experiments are the 10-bus system \textit{case10ba}, the 33-bus system \textit{case33bw}, and the 141-bus system \textit{case141} from M{\sc atpower} \cite{zimmerman_matpower_2011}. For the CLAs, we minimize the $\ell_1$ error with 1000 samples in the first iteration and 4000 additional samples in a sample selection step, and we choose a quadratic output function of voltage magnitude. (See~\cite{BUASON2022} for a discussion on computationally efficient iterative methods for computing CLAs and variants of CLAs that approximate different quantities in order to improve their accuracy.) All power injections vary within 50\% to 150\% of the load demand values given in the M{\sc atpower} files except for \textit{case33bw} where we consider a variant with solar panels \textcolor{black}{at buses 18 and 33. The active power demands at these two buses vary within -200\% to 150\% of the nominal values. Note that a manufacturer can provide actual data regarding the range of power injections from DERs like solar PV.} 

We implement the single-level reformulations of the sensor placement problem in MATLAB using YALMIP~\cite{Lofberg2004} and use Gurobi as a solver with a MIP gap tolerance of $0.5$\%.
The AGD step size is $\epsilon_{AGD} = 2\times10^{-4}$ per unit. The value of $\delta$ (i.e., cost of placing a sensor) in the objective~\eqref{cost_function} is $0.02$. In \textit{case10ba}, \textit{case33bw}, and \textit{case141}, the lower voltage limits are 0.90, 0.91, and 0.92, respectively, and the computation times for the CLAs are 58, 198, and 1415 seconds, respectively.

\subsection{Sensor locations} \label{sec:sensor_locations}
We compare the quality of results and the computation time from the following reformulations: (i)~the KKT formulation,
(ii)~the duality-based bilinear formulation~\eqref{billinear_lower_l} and \eqref{billinear_lower_u},
and (iii)~the MILP formulation with the BVR pre-processing.

\ifthenelse{\boolean{commentOut}}{\begin{table}[t]
\caption{Results showing sensor alarm thresholds and number of false positives from KKT, bilinear, and MILP formulations.}\label{table:sensor_formulation}
\centering
\setlength\tabcolsep{1pt}
\scriptsize
\begin{tabular}{ c c|c|c c c|c c c }
 & &\textbf{KKT}$^\mathsection$ & \multicolumn{3}{c|}{\textbf{Bilinear}} & \multicolumn{3}{c}{\textbf{MILP}} \\
\cline{3-9}
& & \textit{case10ba} 
& \textit{case10ba} & \textit{case33bw} & \textit{case141} 
& \textit{case10ba} & \textit{case33bw} & \textit{case141} \\
\hline \hline
Computation & \multicolumn{1}{ |c| }{Optimality} & 26.7 & 1.96 & 4.47 & 46.52 & 1.54 & 2.87 & 22.95 \\ \cline{2-9}
time [\si{\second}]& \multicolumn{1}{ |c| }{AGD} & \NA & 0.11 & 0.31 & 18.3 &  \NA & 0.43 & 13.8 \\ \hline
\multicolumn{2}{c|}{Sensor location(s)} & 10 & 10 & 14, 15, 17, 31 & 79, 80, 82, 85 & 10 & 14, 30 & 80, 86  \\ \hline
\multicolumn{2}{c|}{\multirow{2}{*}{Sensor threshold(s)}}        & \multirow{2}{*}{0.9} & \multirow{2}{*}{0.9017} & 0.91, 0.91, & 0.92, 0.9213, & \multirow{2}{*}{0.9} & 0.9195 & 0.929  \\ 
& & & & 0.9118, 0.9126 & 0.93, 0.93 & & 0.9185 & 0.9295  \\ \hdashline
\multicolumn{2}{c|}{\multirow{2}{*}{with AGD}}       & \multirow{2}{*}{\NA} & \multirow{2}{*}{0.9} & 0.91, 0.91, & 0.92, 0.9212, & \multirow{2}{*}{\NA} & 0.9163 &  0.9213 \\ 
&   &   & & 0.9107, 0.9112 & 0.9218, 0.9201 &  & 0.9161 & 0.9201  \\ \hline
\multicolumn{2}{c|}{\# feasible points} & 7317 & 7317 & 9753 & 9955 & 7317 & 9753 & 9955   \\ \hline
\multicolumn{2}{c|}{\% false positive(s)} & 0\% & 4.01\% & 1.91\% & 72.07\% & 0\% & 7.64\% & 66.64\%  \\ \hdashline
\multicolumn{2}{c|}{with AGD} & \NA & 0\% & 0.24\% & 0.03\% & \NA & 1.34\% & 0.01\% \\ \hline
\multicolumn{2}{c|}{\% false negatives} & 0\% & 0\% & 0\% & 0\% & 0\% & 0\% & 0\%   \\ \hline \hline
\multicolumn{9}{l}{%
  \begin{minipage}{13.5cm}%
    \footnotesize $^\mathsection$The solver does not find a solution to \textit{case33bw} and \textit{case141} within 55000 seconds.
  \end{minipage}%
}\\
\end{tabular}
\end{table}}
{\begin{table}[t]
\caption{Results showing sensor alarm thresholds and number of false positives from KKT, bilinear, and MILP formulations.}\label{table:sensor_formulation}
\centering
\setlength\tabcolsep{2pt}
\scriptsize
\begin{tabular}{ c c|c|c c c|c c c|c c c }
 & &\textbf{KKT}$^\mathsection$ & \multicolumn{3}{c|}{\textbf{Bilinear}} & \multicolumn{3}{c|}{\textbf{MILP w/o BVR}} & \multicolumn{3}{c}{\textbf{MILP}} \\
\cline{3-12}
& & \textit{case10ba} 
& \textit{case10ba} & \textit{case33bw} & \textit{case141} & \textit{case10ba} & \textit{case33bw} & \textit{case141}
& \textit{case10ba} & \textit{case33bw} & \textit{case141} \\
\hline \hline
Computation & \multicolumn{1}{ |c| }{Optimality} & 26.7 & 1.96 & 4.47 & 46.52 & 1.95 & 3.19 & 69.6 & 1.54 & 2.87 & 22.95 \\ \cline{2-12}
time [\si{\second}]& \multicolumn{1}{ |c| }{AGD} & \NA & 0.11 & 0.31 & 18.3 & \NA & 0.83 & 16.5 &  \NA & 0.43 & 13.8 \\ \hline

\multicolumn{2}{c|}{Sensor location(s)} & 10 & 10 & 14, 15, 17, 31 & 79, 80, 82, 85 & 10 & 14, 31 & 76, 80, 86 & 10 & 14, 30 & 80, 86  \\ \hline
\multicolumn{2}{c|}{\multirow{2}{*}{Sensor threshold(s)}}        & \multirow{2}{*}{0.9} & \multirow{2}{*}{0.9017} & 0.91, 0.91, & 0.92, 0.9213, & \multirow{2}{*}{0.9}  & 0.9195 & 0.92, 0.9295, & \multirow{2}{*}{0.9} & 0.9195 & 0.929  \\ 
& & & & 0.9118, 0.9126 & 0.93, 0.93 &  & 0.9195  & 0.9295 & & 0.9185 & 0.9295  \\ \hdashline
\multicolumn{2}{c|}{\multirow{2}{*}{with AGD}}       & \multirow{2}{*}{\NA} & \multirow{2}{*}{0.9} & 0.91, 0.91, & 0.92, 0.9212, & \multirow{2}{*}{\NA} & 0.9158 & 0.92, 0.9211, & \multirow{2}{*}{\NA} & 0.9163 &  0.9213 \\ 
&   &   & & 0.9107, 0.9112 & 0.9218, 0.9201 &  & 0.9125  & 0.9201 &  & 0.9161 & 0.9201  \\ \hline
\multicolumn{2}{c|}{\# feasible points} & 7317 & 7317 & 9753 & 9955 & 7317 & 9753 & 9955 & 7317 & 9753 & 9955   \\ \hline
\multicolumn{2}{c|}{\% false positive(s)} & 0\% & 4.01\% & 1.91\% & 72.07\% & 0\% & 18.97\% & 66.64\% & 0\% & 7.64\% & 66.64\%  \\ \hdashline
\multicolumn{2}{c|}{with AGD} & \NA & 0\% & 0.24\% & 0.03\% & \NA & 1.55\% & 0.01\% & \NA & 1.34\% & 0.01\% \\ \hline
\multicolumn{2}{c|}{\% false negatives} & 0\% & 0\% & 0\% & 0\% & 0\% & 0\% & 0\% & 0\% & 0\% & 0\%   \\ \hline \hline
\multicolumn{10}{l}{%
  \begin{minipage}{13.5cm}%
    \footnotesize $^\mathsection$The solver does not find a solution to \textit{case33bw} and \textit{case141} within 55000 seconds.
  \end{minipage}%
}\\
\end{tabular}
\end{table}}

\begin{figure}[th!] 
    \centering
  \subfloat[\textit{case10}]{\label{fig:bar_case10}\includegraphics[trim=3.5cm 8cm 4.5cm 8.5cm, clip, width=0.5\linewidth]{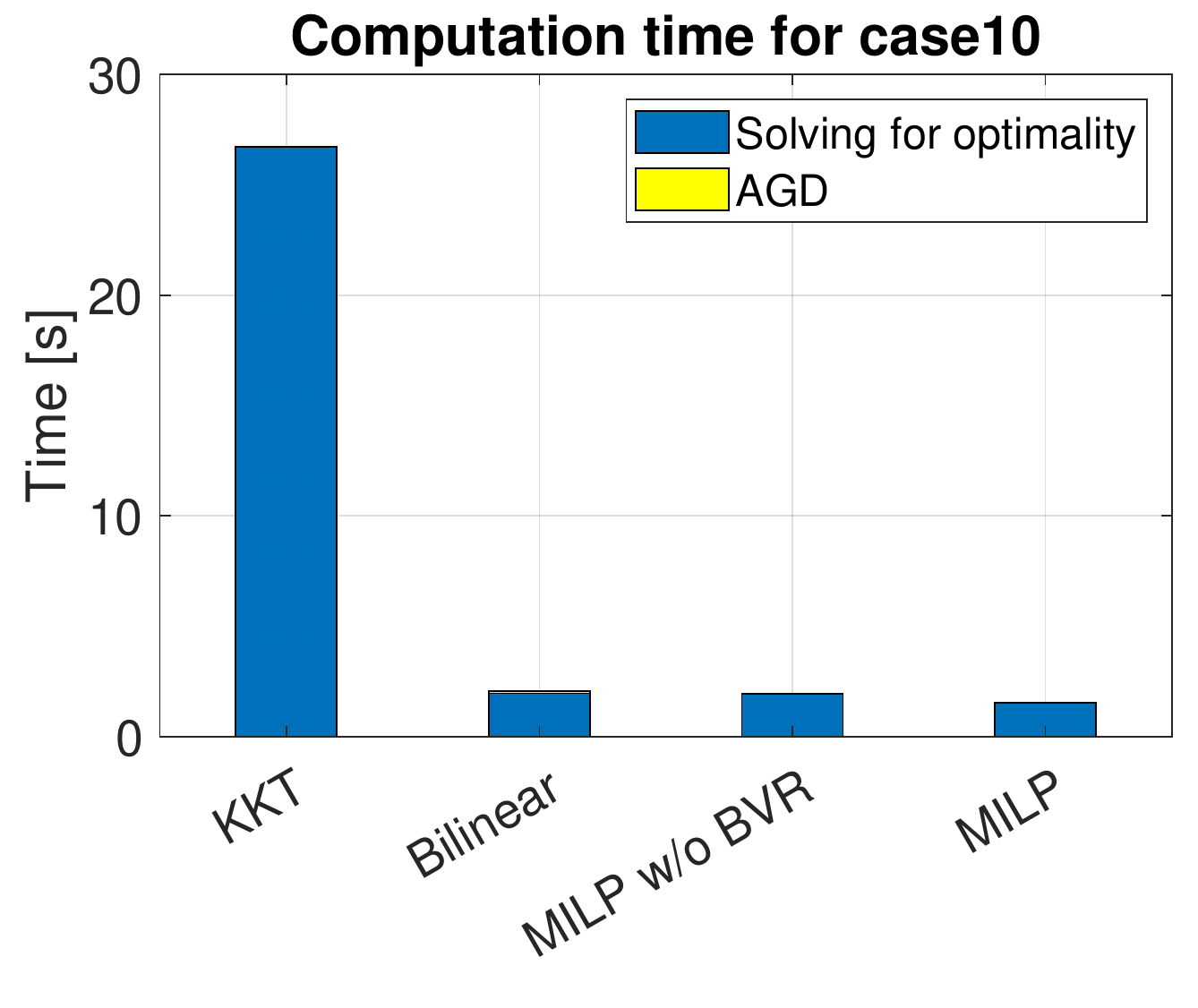}}
    \hfill
  \subfloat[\textit{case33bw}]{\label{fig:bar_case33}\includegraphics[trim=3.5cm 8cm 4.5cm 8.5cm, clip, width=0.5\linewidth]{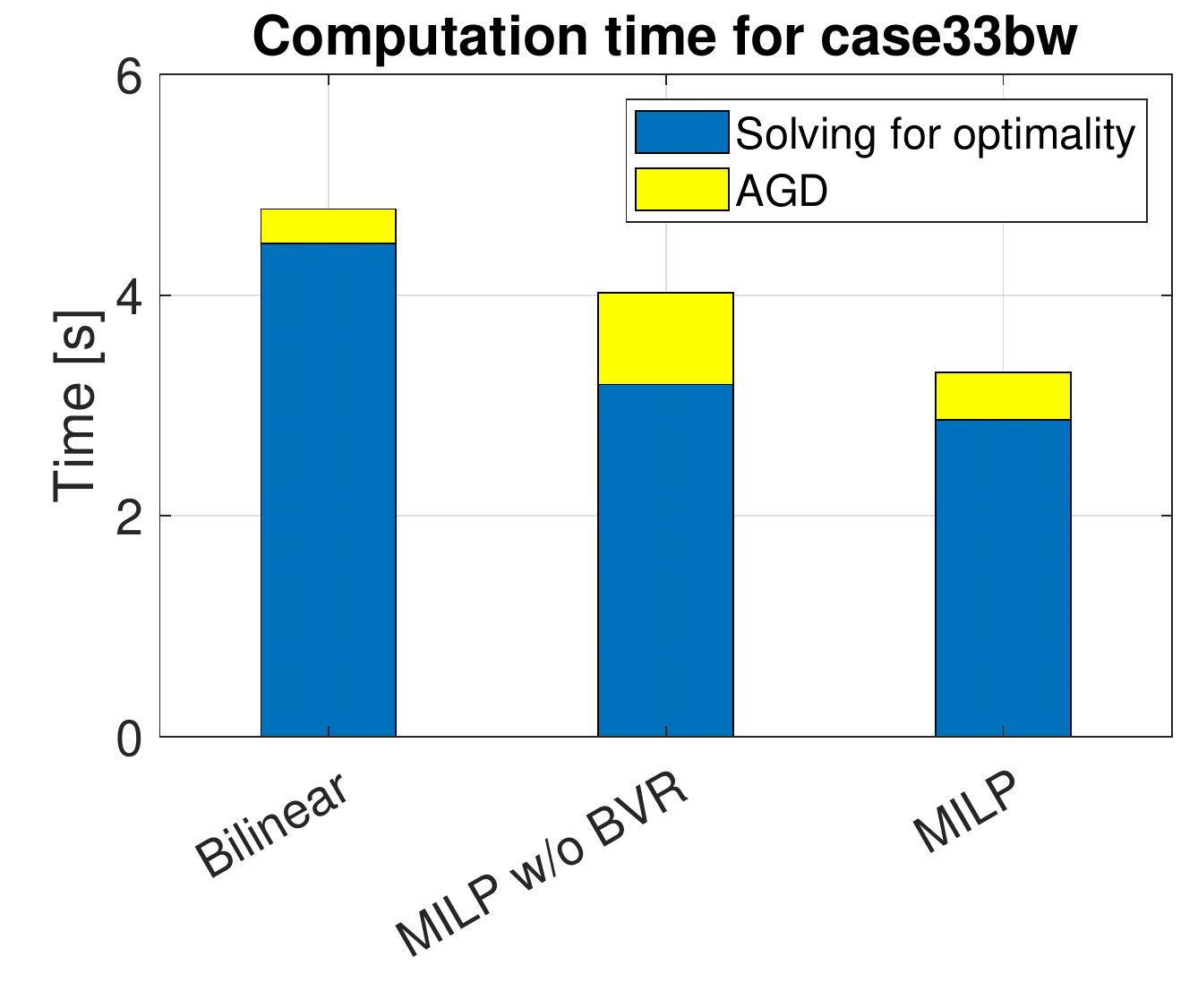}}
  \hfill
  
    \subfloat[\textit{case141}]{\label{fig:bar_case141}\includegraphics[trim=3.5cm 8cm 4.5cm 8.5cm, clip, width=0.5\linewidth]{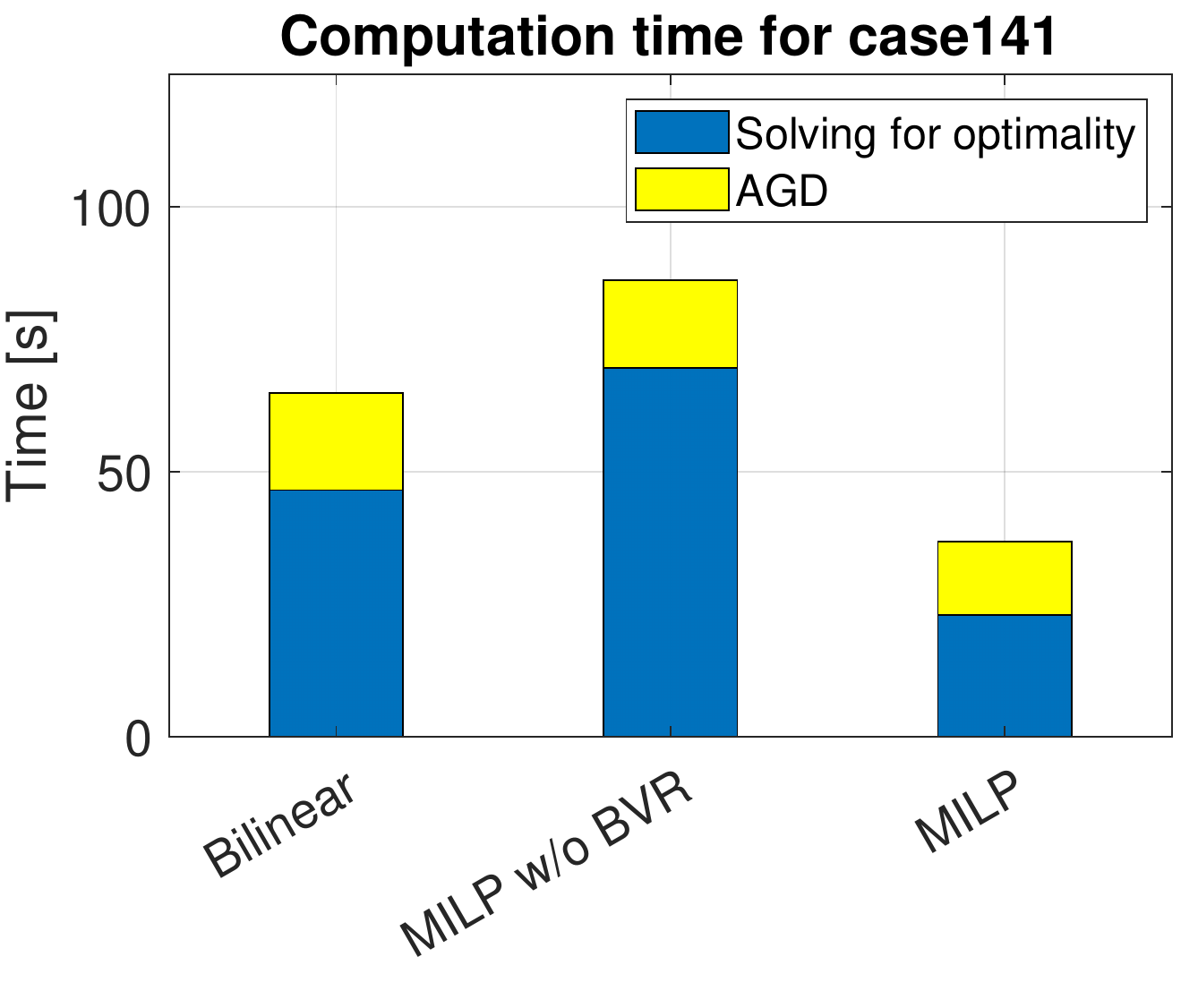}}
   \caption{Bar graphs showing the computation times for solving various problem formulations and executing the approximate gradient descent method.}
  \label{fig:computation times} 
\end{figure}

The first test case is the 10-bus system \textit{case10ba}, a simple single-branch network. We consider a variant where the nominal loads are $60\%$ of the values in the M{\sc atpower} file. The results from each formulation place a sensor at the end of the branch (furthest bus from the substation) with an alarm threshold of $0.9$ per unit (at the voltage limit). Fig.~\ref{fig:bar_case10} compares computation times from the three formulations. The the KKT formulation takes 26.7 seconds while the bilinear and MILP formulations take 1.96 and 1.54 seconds, respectively. Since the sensor threshold for the KKT and MILP formulations is at the voltage limit, AGD is not needed. Conversely, the bilinear formulation gives a higher alarm threshold. As a result, the AGD method is applied as a post-processing step to achieve the lowest possible threshold without introducing false alarms. The number of false positives reduces from $5.48\%$ to $0\%$. Executing the AGD method takes 0.11 seconds.

The second test case is the 33-bus system \textit{case33bw}, which has multiple branches. \textcolor{black}{In this example, we demonstrate the efficacy of our approach in handling a system with complex components through the implementation of volt-VAR control, which represents smarter inverter behavior (whose characteristics are described in~\cite{osti2016}). To incorporate the behavior of volt-VAR control, we enhance the power flow solver used to compute the CLAs by integrating an additional fixed-point iterative method.} Table~\ref{table:sensor_formulation} shows the computation times for the bilinear and the two MILP formulations. We exclude the computation time for the KKT formulation since the solver fails to find even a feasible (but potentially suboptimal) point within 55000 seconds (15 hours). Our final test case is the 141-bus system \textit{case141}. Similar to the 33-bus system, the solver could not find the optimal solution for the KKT formulation within a time limit of 15 hours. \textcolor{black}{It is evident the KKT formulation is intractable.} Table~\ref{table:sensor_formulation} again shows the results for this test case, and Figs.~\ref{fig:bar_case33} and~\ref{fig:bar_case141} compare the computation times for the bilinear and MILP formulations.

Table~\ref{table:sensor_formulation} shows both the computation times and the results of randomly drawing sampled power injections within the specified range of variability, computing the associated voltages by solving the power flow equations, and finding the number of false positive alarms (i.e., the voltage at a bus with a sensor is outside the sensor's threshold but there are no voltage violations in the system). The results for the 33-bus and 141-bus test cases given in Table~\ref{table:sensor_formulation} illustrate the performance of the proposed reformulations. Whereas the KKT formulation is computationally intractable, our proposed reformulations find solutions within approximately one minute, where the MILP formulation with the BVR method typically exhibits the fastest performance. The solutions to the reformulated problems place a small number of sensors (two to four sensors in systems with an order of magnitude or more buses). No solutions suffer from false negatives since all samples where there is a voltage violation trigger an alarm. There are a number of false alarms prior to applying the AGD that after its application decrease dramatically to a small fraction of the total number of samples ($1.34\%$ and $0.01\%$ in the 33-bus and the 141-bus systems, respectively). These observations suggest that our sensor placement formulations provide a computationally efficient method for identifying a small number of sensor locations and associated alarm thresholds that reliably identify voltage constraint violations with no false negatives (missed alarms) and few false positives (spurious alarms).

\subsection{Multiple configurations} \label{sub:configuration_sim}
\begin{figure}[ht!]
	\centering 
	\includegraphics[trim=0.5cm 0.5cm 0.4cm 0.4cm, clip, width=0.4\linewidth]{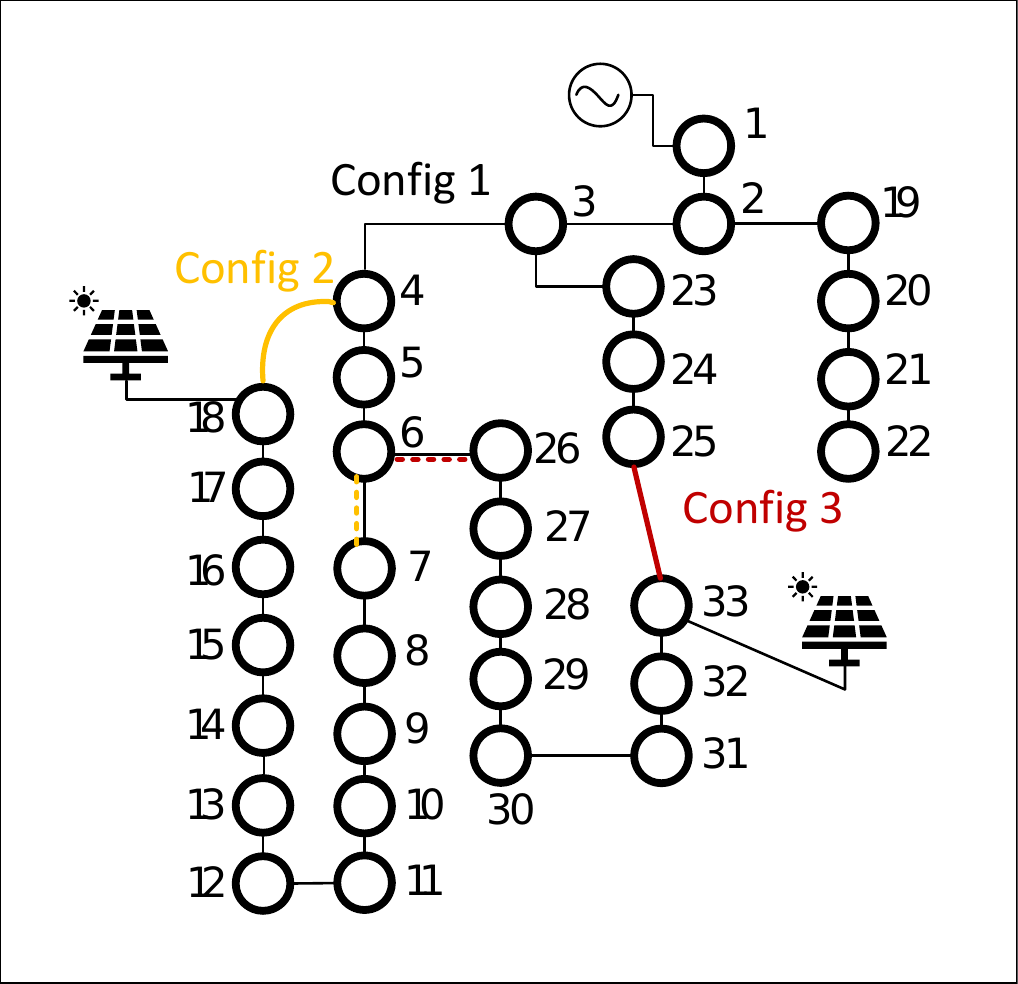} 
	\caption{The \textit{case33bw} system with three different network configurations. A dash line means that line is removed when a solid line with the same color is added such that the system has a radial topology in each configuration.}
	\label{fig:case33_allConfig}
	\vspace{-0.5em}
\end{figure}

The previous results described the sensor placements for the \textit{case10ba}, \textit{case33bw}, and \textit{case141} systems in their nominal network topologies. We next demonstrate the effectiveness of our problem reformulations for variants of these systems with multiple network configurations. We consider a variant of the \textit{case33bw} system with three distinct network configurations and two solar PV generators installed at buses 18 and 33, as an illustrative example. Other network configurations are not included from this study, as they either exhibit no violations or yield identical optimal solutions.
The first configuration is the nominal topology given in the M{\sc atpower} version of the test case. In the second configuration, the line connecting buses 6 and 7 is removed and a new line connecting buses 4 and 18 is added. The third configuration removes the line connecting buses 6 and 26 and adds a new line connecting buses 25 and 33. All network configurations are displayed in Fig.~\ref{fig:case33_allConfig}.

Table~\ref{table:case33_sensor_same_location} shows the results from using the bilinear and MILP formulations to solve the multiple-configuration problem for this case. The results generally mirror those from the single-network-configuration test cases shown earlier in that computation times are still reasonable (approximately a factor of four larger) and there are no false negatives and a small number of false positives after applying the AGD method.

\begin{table}[t]
\caption{Results for the \textit{case33bw} system with three network configurations.} \label{table:case33_sensor_same_location}
\centering
\setlength\tabcolsep{1.5pt}
\scriptsize
\begin{tabular}{ c c|c c c| c c c }
 & & \multicolumn{3}{c|}{\textbf{Bilinear}} & \multicolumn{3}{c}{\textbf{MILP}} \\
\cline{3-8}
& & \textbf{Config 1} & \textbf{Config 2} & \textbf{Config 3} & \textbf{Config 1} & \textbf{Config 2} & \textbf{Config 3}\\
\hline \hline
Computation & \multicolumn{1}{ |c| }{Optimality} & \multicolumn{3}{c|}{\longdash[15] 20.1 \longdash[15]} &  \multicolumn{3}{c}{\longdash[15] 7.93 \longdash[15]}   \\ \cline{2-8}
time [\si{\second}]& \multicolumn{1}{ |c| }{AGD} & 0.53 & 0.55 & 0.10 & 0.86 & 0.93 & 0.40  \\ \hline
\multicolumn{2}{c|}{Sensor location(s)} & \multicolumn{3}{c|}{\longdash[10] 8, 14, 26, 33 \longdash[10]} &  \multicolumn{3}{c}{\longdash[15] 9, 14, 30 \longdash[15]}  \\ \hline
\multicolumn{2}{c|}{\multirow{2}{*}{Sensor thresholds}}  & 0.91, 0.919, & 0.9126, 0.91,  & 0.91, 0.91, & 0.91, 0.919,     & 0.9185, 0.91, & 0.91, 0.91, \\
& & 0.91, 0.9113 & 0.91, 0.9111 & 0.9106, 0.9137 & 0.919 & 0.919 & 0.9165 \\ \hdashline
\multicolumn{2}{c|}{\multirow{2}{*}{with AGD}}  & 0.91, 0.9167, & 0.9117, 0.91, & 0.91, 0.91,    & 0.91, 0.9167,     & 0.9166, 0.91, & 0.91, 0.91,  \\
& & 0.91, 0.9104 & 0.91, 0.9109 & 0.9101, 0.9136   & 0.9168 & 0.9189 & 0.9151 \\ \hline
\multicolumn{2}{c|}{\# feasible points} & 9560 & 8292 & 9112 & 9560 & 8292 & 9112  \\ \hline
\multicolumn{2}{c|}{\% false positives} & 6.19\% & 4.86\% & 1.28\% & 9.19\% & 12.38\% & 6.55\%  \\ \hdashline
\multicolumn{2}{c|}{with AGD} & 1.86\% & 1.21\% & 0.05\% & 2.87\% & 3.38\% & 2.38\% \\ \hline
\multicolumn{2}{c|}{\% false negatives} & 0\% & 0\% & 0\% & 0\% & 0\% & 0\%  \\ \hline \hline
\end{tabular}
\end{table}

We note that some configurations may not need to utilize all available sensors. To show this, we describe an experiment that considers each configuration separately. 
In this experiment, we compare the number of sensors and the locations of the sensors with those in the previous experiment. As Table~\ref{table:case33_sensor_different_location} shows, configurations 1 and 2 require only two sensors while configuration 3 requires only one sensor as opposed to three-sensor solution obtained from the multiple-configuration problem. This demonstrates the need to jointly consider network topologies in one problem for such situations.

\begin{table}[h]
\caption{Results for \emph{case33bw} with three network configurations where sensor locations are not necessary the same.}\label{table:case33_sensor_different_location}
\centering
\setlength\tabcolsep{3pt}
\small
\begin{tabular}{ c c|c|c|c }
& & \textbf{Config 1} & \textbf{Config 2} & \textbf{Config 3} \\
\hline \hline
\multicolumn{2}{c|}{Sensor location(s)} & 14, 30 & 9, 31 & 30  \\ \hline
\multicolumn{2}{c|}{Sensor}       & 0.9195     & 0.9185 & \multirow{2}{*}{0.9185}  \\ 
\multicolumn{2}{c|}{threshold(s)}   & 0.9190 & 0.9185 \\ \hdashline
\multicolumn{2}{c|}{\multirow{2}{*}{with AGD}}     & 0.9167     & 0.9164 & \multirow{2}{*}{0.9151}  \\
& & 0.9168 & 0.9186 & \\ \hline
\multicolumn{2}{c|}{\# feasible points} & 9560 & 8292 & 9112  \\ \hline
\multicolumn{2}{c|}{\% false positives} & 10.48\% & 12.45\% & 14.53\%  \\ \hdashline
\multicolumn{2}{c|}{with AGD} & 2.89\% & 3.06\% &  2.38\% \\ \hline
\multicolumn{2}{c|}{\% false negatives} & 0\% & 0\% & 0\%   \\ \hline \hline
\end{tabular} \label{table:case33_eachconfig}
\end{table}

\ifthenelse{\boolean{commentOut}}{}{\subsection{Comparison to a heuristic sensor placement technique} \label{sec:simulation_heuristic}
We next demonstrate the performance of the heuristic sensor placement technique described in Section~\ref{sec:heuristic} in the context of multiple network configurations for the three-configuration variant of the test case \textit{case33bw}. We study the effects of two versions of this heuristic: (i)~place sensors at the end of all branches based on configuration 1 and (ii)~place sensors at the ends of branches considering all configurations. The results are shown in Table~\ref{table:heuristic}. For the first version of this heuristic, we place four sensors. This technique works well in configuration~1; however, it introduces false negatives (failures to alarm in cases with voltage limit violations) in approximately 10\% to 20\% of the power injection samples for configurations~2 and~3 since the sensors are instead located in the middle of some branches and thus do not capture all possible violations. To reduce the number of false negatives, we next consider sensor locations based on all configurations, i.e., the second version of the heuristic. This results in six sensors being placed and only one occurrence of a false negative. Comparing to our approach (refer to Table~\ref{table:case33_sensor_same_location}), we only need three sensors (as opposed to six sensors) to detect all violations. This shows the necessity of using an optimization formulation to obtain \emph{sparse} sensor placement solutions, since a naive approach where sensors are placed at all buses will always avoid false positives.

\begin{table*}[h]
\caption{Results for \emph{case33bw} with three network configurations using two variants of the heuristic technique} \label{table:heuristic}
\centering
\setlength\tabcolsep{3pt}
\small
\begin{tabular}{ c|c|c c c }
\textbf{Case} & & \textbf{Config 1} & \textbf{Config 2} & \textbf{Config 3} \\
\hline \hline
\multirow{5}{*}{\begin{turn}{90}\textbf{version 1}\end{turn}} & Sensor location(s) & \multicolumn{3}{c}{\longdash[10] 18, 22, 25, 33 \longdash[10]} \\ \cline{2-5}
& Sensor thresholds  & \multicolumn{3}{c}{\longdash[5] 0.91, 0.91, 0.91, 0.91 \longdash[5]}\\ \cline{2-5}
& \# feasible points & 9560 & 8292 & 9112 \\ \cline{2-5}
& \% false positives & 0\% & 0\% & 0\%  \\ \cline{2-5}
& \% false negatives & 0\% & 16.26\% & 8.85\%  \\ \hline\hline

\multirow{5}{*}{\begin{turn}{90}\textbf{version 2}\end{turn}} & Sensor location(s) & \multicolumn{3}{c}{\longdash[5] 7, 18, 22, 25, 26, 33 \longdash[5]} \\ \cline{2-5}
& Sensor thresholds  & \multicolumn{3}{c}{0.91, 0.91, 0.91, 0.91, 0.91, 0.91} \\ \cline{2-5}
& \# feasible points & 9560 & 8292 & 9112 \\ \cline{2-5}
& \% false positives & 0\% & 0\% & 0\%  \\ \cline{2-5}
& \% false negatives & 0.01\% & 0\% & 0\%  \\ \hline \hline
\end{tabular}
\end{table*}}

\section{Conclusion} \label{sec:conclusion}

This paper has formulated a bilevel optimization problem that seeks to minimize the number of sensors needed to detect violations of voltage magnitude limits in an electric distribution system. We first addressed the power flow nonlinearities in the lower-level problem via previously developed conservative linear approximations of the power flow equations. 
To handle computational challenges from the bilevel nature, we exploited the structure of the problem to obtain single-level mixed-integer programming formulations, thereby avoiding the introduction of unnecessary additional discrete variables. We also developed a mixed-integer-linear programming (MILP) formulation by discretizing the sensor thresholds. Our MILP reformulation proves tractable across all test cases, a notable advantage over the traditional KKT reformulation, which only remains tractable for the 10-bus system. Furthermore, the MILP formulation yields a reduced number of sensors compared to the bilinear formulation. Additionally, we extended these reformulations to accommodate multiple network topologies, wherein several lines can be opened or closed without any nodes being isolated. Our proposed sensor placement reformulations significantly reduce computation time compared to standard techniques.

Furthermore, we developed a post-processing technique to minimize false alarms using an approximate gradient descent method. The integration of this technique with our bilevel problem reformulation ensures the computation of sensor locations and alarm thresholds that yield minimal false alarms and no missed alarms. This was validated through numerical testing conducted via out-of-sample testing.

For enhanced computational efficiency, we may further expedite computations by considering sensor placements solely at locations where voltage violations occurred, utilizing insights from sample-based CLAs. Additionally, employing a constraint generation approach enables us to explicitly formulate constraints associated only with lower-level problems where violations were observed in the samples used for CLA computation. Essentially, leveraging samples from CLA computation minimizes the necessity for explicit modeling of variables and constraints in the bilevel sensor placement problem. Subsequently, conducting out-of-sample testing, as demonstrated in our numerical analyses, can validate the efficacy of these approaches in further enhancing tractability.

In our future work, we seek to identify where the violations occur using the information obtained from CLAs and solutions from the sensor placement problem. We intend to use the sensor locations and thresholds resulting from the proposed formulations to design corrective control actions, including voltage controls, that ensure all voltages remain within safety limits. These actions may also involve adjustments to generation levels, switching devices, or control of reactive power sources.

Expanding our research scope, we aim to work on real-time operations, focusing particularly on optimizing the utilization of the limited communication bandwidth accessible at smart meters. Our objective is to devise strategies to efficiently extract the most relevant real-time information from a network of sensors, therefore enhancing the overall operational effectiveness. Additionally, we intend to address this challenge while considering potential cyber threats, such as false data injection attacks~\cite{Liang2017}.

Additionally, it is worth noting that prior research has demonstrated the tractability of extending the CLA framework to unbalanced three-phase systems \cite{aquino2023managing}. We intend to explore the extension of our work to three-phase networks in future work.

\section*{Acknowledgement}
Support from NSF \#2023140 and PSERC \#T-64 (P.~Buason and D.K.~Molzahn), NSF Graduate Research Fellowship Program under Grant No. DGE-1650044 (S. Talkington), and DOE Office of Electricity Advanced Grid Modeling Program (S.~Misra).

\bibliographystyle{IEEEtran}
\bibliography{reference.bib}

\newpage

\appendix 

\section{Reformulation using KKT constraints} \label{sec:kkt}
With a linear lower-level problem, we can apply standard techniques for reformulating the bilevel problem~\eqref{full_problem} with CLAs as a (single-level) mixed-integer linear program (MILP). These techniques replace the lower-level problem~\eqref{lower_level_objective_cla} with its KKT conditions that are both necessary and sufficient for optimality of this problem~\cite{dempe2012} and also apply McCormick envelopes~\cite{mccormick1976} to convert the bilinear product of the continuous and discrete variables in~\eqref{sens_limit_upper_full} to an equivalent linear form. The resulting single-level problem still involves bilinear constraints associated with the complementarity conditions. These bilinear constraints are traditionally addressed using binary variables in a ``Big-M'' formulation. Commercial MILP solvers are applicable to this traditional reformulation, which we denote throughout the paper as the ``KKT formulation''. This formulation is obtained by defining the lower-level coupling quantities $\llevl_i$ and $\ulevl_i$ using the KKT conditions given below:
\begin{subequations} \label{eq:kkt_reformulation}
\begin{align} 
\llevl_i = \ &\underline{a}_{i,0} +  \underline{\bm{a}}_{i,1}^{T} \begin{pmatrix}
        \mathbf{P} \\ \mathbf{Q} 
        \end{pmatrix}^i \quad \bigg(\ulevl_i = \ \overline{a}_{i,0} +  \overline{\bm{a}}_{i,1}^{T} \begin{pmatrix}
        \mathbf{P} \\ \mathbf{Q} 
        \end{pmatrix}^i\bigg),  \label{lower_level_objective_kkt} \\
        (&\forall j \in N_{PQ} \setminus \{i\}) \nonumber \\
&\underline{a}_{i,1} - \sum_{k\in\mathcal{N}_{PQ}} \lambda_k^i \overline{\bm{a}}_{k,1} + \sum_{k\in\mathcal{N}_{PQ}} \mu_k^i \underline{\bm{a}}_{k,1}  \nonumber \\
        & \qquad - \sum_{k\in\mathcal{N}_{PQ}} \gamma_k^i e_k + \sum_{k\in\mathcal{N}_{PQ}} \pi_k^i e_k = 0 \nonumber \\
(\text{for} \; \ulevl_i: \ & \overline{a}_{i,1} - \sum_{k\in\mathcal{N}_{PQ}} \lambda_k^i \overline{\bm{a}}_{k,1} + \sum_{k\in\mathcal{N}_{PQ}} \mu_k^i \underline{\bm{a}}_{k,1}  \nonumber \\
        &\qquad  - \sum_{k\in\mathcal{N}_{PQ}} \gamma_k^i e_k + \sum_{k\in\mathcal{N}_{PQ}} \pi_k^i e_k = 0), \label{stationarity_kkt} \\
    &\overline{a}_{j,0} +  \overline{\bm{a}}_{j,1}^{T} \begin{pmatrix}
            \mathbf{P} \\ \mathbf{Q} 
            \end{pmatrix}^i \geq \undertilde{U}_j, \label{primal_kkt_1}\\
    &\underline{a}_{j,0} +  \underline{\bm{a}}_{j,1}^{T} \begin{pmatrix}
            \mathbf{P} \\ \mathbf{Q} 
            \end{pmatrix}^i \leq \widetilde{U}_j, \label{primal_kkt_2} \\
&\begin{pmatrix}
        \mathbf{P} \\ \mathbf{Q} 
        \end{pmatrix}^{\text{min}} \leq \begin{pmatrix}
        \mathbf{P} \\ \mathbf{Q} 
        \end{pmatrix}^i  \leq  \begin{pmatrix}
        \mathbf{P} \\ \mathbf{Q} 
        \end{pmatrix}^{\text{max}}, \label{primal_kkt_3}  \\
&\lambda_j^i \cdot \left(\undertilde{U}_j - \overline{a}_{j,0} -  \overline{\bm{a}}_{j,1}^{T} \begin{pmatrix}
        \mathbf{P} \\ \mathbf{Q} 
        \end{pmatrix}^i\right) = 0, \label{comp_slack_kkt_1} \\
&\mu_j^i \cdot \left(-\widetilde{U}_j + \underline{a}_{j,0} +  \underline{\bm{a}}_{j,1}^{T} \begin{pmatrix}
        \mathbf{P} \\ \mathbf{Q} 
        \end{pmatrix}^i\right) = 0, \label{comp_slack_kkt_2} \\
&\boldsymbol{\gamma}^i \odot \left(\begin{pmatrix}
        \mathbf{P} \\ \mathbf{Q} 
        \end{pmatrix}^{\text{min}} - \begin{pmatrix}
        \mathbf{P} \\ \mathbf{Q} 
        \end{pmatrix}^i\right) = \bm{0}, \label{comp_slack_kkt_3} \\
&\boldsymbol{\pi}^i \odot \left(\begin{pmatrix}
        \mathbf{P} \\ \mathbf{Q} 
        \end{pmatrix}^i - \begin{pmatrix}
        \mathbf{P} \\ \mathbf{Q} 
        \end{pmatrix}^{\text{max}}\right) = \bm{0}, \label{comp_slack_kkt_4} \\
&\boldsymbol{\gamma}^i, \boldsymbol{\pi}^i \geq \mathbf{0}; \lambda_j^i, \mu_j^i \geq 0, \; \forall j \in N_{PQ} \setminus \{i\}, \label{dual_kkt}
\end{align}
\end{subequations}
where the operator $\odot$ is the element-wise multiplication; $e_i$ is the $i^{\text{th}}$ column of the identity matrix; $\boldsymbol{\lambda}$, $\boldsymbol{\mu}$, $\boldsymbol{\gamma}~:= (\gamma_1, \gamma_2, \dots, \gamma_{2m})^T$, and $\boldsymbol{\pi}~:= (\pi_1, \pi_2, \dots, \pi_{2m})^T$ are dual variables associated with the voltage and power injection bounds. Note that the solution to the set of equations in $\llevl_i$ is completely decoupled from that in $\ulevl_i$. Equations \eqref{stationarity_kkt}--\eqref{comp_slack_kkt_4} are the KKT conditions of the lower-level problem. Equation \eqref{stationarity_kkt} is the stationarity condition. Equations \eqref{primal_kkt_1}--\eqref{primal_kkt_3} are the primal feasibility conditions. The complementary slackness conditions are \eqref{comp_slack_kkt_1}--\eqref{comp_slack_kkt_4} and the dual feasibility condition is \eqref{dual_kkt}. Observe that the complementary slackness conditions give rise to nonlinear functions due to the multiplication of the dual variables $\boldsymbol{\lambda}$, $\boldsymbol{\mu}$, $\boldsymbol{\gamma}$, and $\boldsymbol{\pi}$ with the primal variables $\mathbf{P}$ and $\mathbf{Q}$. To handle these nonlinearities, traditional methods for bilevel optimization replace these products using additional binary variables and a big-M reformulation. This requires bounds on the dual variables that are difficult to determine, and bad choices for these bounds can result in either infeasibility or poor computational performance~\cite{PINEDA2019}.

\section{Bilinear to mixed-integer linear programming}\label{subsec:milp_reformulation}
The bilinear formulation can be further converted into an MILP by discretizing the continuous-valued sensor thresholds. This formulation has the advantage of being within the scope of a larger range of mixed-integer programming solvers since not all of them can handle billinear forms. We partition the sensor threshold ranges into $d$ discrete steps with size $\epsilon$ and define the vectors of threshold variables, $\undertilde{\mathbf{U}}$ and $\widetilde{\mathbf{U}}$, as 
\begin{equation}\label{discretized_sensor_thresholds}
\undertilde{\mathbf{U}} = \eta^T \undertilde{v}_t \quad (\widetilde{\mathbf{U}} = \eta^T \widetilde{v}_t), 
\end{equation}
where
\begin{subequations}
\begin{align}
&\eta = 
\begin{bmatrix}
\eta_{0,1} & \eta_{0,2} & \cdots & \eta_{0,n} \\
\eta_{1,1} & \eta_{1,2} & \cdots & \eta_{1,n} \\
\vdots  & \vdots  & \ddots & \vdots  \\
\eta_{d,1} & \eta_{d,2} & \cdots & \eta_{d,n} 
\end{bmatrix}, \label{eq:eta}\\
&\undertilde{v}_t = 
\begin{bmatrix}
0, 
V^{\text{min}}, 
V^{\text{min}} + \epsilon, \ldots, V^{\text{min}} + (d - 1)\epsilon
\end{bmatrix}^T \nonumber \\
(&\widetilde{v}_t = 
\begin{bmatrix}
0, 
V^{\text{max}} - \epsilon, V^{\text{max}} - 2\epsilon, \ldots,  
V^{\text{max}}
\end{bmatrix}^T), \label{eq:threshold_discrete} \\
&\sum_{i = 1}^{d} \eta_{i,k} = 1, \quad \forall k \in \mathcal{N}_{PQ} \label{eq:sum_threshold_one}.
\end{align}
\end{subequations}

Note that this discretization exploits the fact that any sensor threshold will necessarily be above the lower voltage limit $V_i^{\text{min}}$ and below the upper voltage limit $V_i^{\text{max}}$. Equations~\eqref{eq:eta}--\eqref{eq:sum_threshold_one} imply that when $\eta_{0,i} = 0$, no sensor is placed (i.e., $s_i = 0$). Using this discretization, the constraints \eqref{lower_level_objective_dual_single_l} and \eqref{lower_level_objective_dual_single_u} now contain bilinear products of binary variables. These products can be equivalently transformed into a mixed-integer \emph{linear} (as opposed to bilinear) programming formulation using McCormick Envelopes~\cite{mccormick1976}. With McCormick Envelopes and discrete sensor thresholds, the problems~\eqref{billinear_lower_l} and \eqref{billinear_lower_u} become a MILP that can be computed using any MILP solver. We refer to the reformulation of the lower-level problems~\eqref{billinear_lower_l} and \eqref{billinear_lower_u} using the discretization~\eqref{discretized_sensor_thresholds} as the ``MILP formulation''.

To further improve tractability, we can remove unnecessary binary variables by inspecting data from the samples of power injections used to compute the conservative linear approximations of the power flow equations. Let $b$ be a bus where the voltage magnitude never reaches the highest discretized sensor threshold value (i.e., $V^{\text{min}} + (d  - 1)\epsilon$) in any of the sampled power injections. Given a sufficiently comprehensive sampling of the range of possible power injections, we can then simplify the discretized representation of the sensor alarm threshold as:
\begin{subequations}
\begin{align}
    &\undertilde{U}_b = 0 \cdot \eta_{0,b} + V^{\text{min}} \cdot \eta_{1,b}, \\
	&\eta_{0,b} + \eta_{1,b} = 1.
\end{align}
\end{subequations}

A similar simplification can be used for the upper sensor thresholds. This pre-screening thus eliminates binary variables associated with sensor thresholds at buses that will never violate their voltage limits. We henceforth call this data-driven simplification technique ``binary variable removal'' (BVR).

\section{Comparisons of each formulation} \label{sec:comparison}
The previous subsections present several problem reformulations that convert the bilevel sensor placement problem~\eqref{full_problem} into various single-level problems that can be solved with mixed-integer solvers like Gurobi. Each reformulation has different computational characteristics and yields solutions with differing accuracy. We next compare the KKT formulation~\eqref{eq:kkt_reformulation} described in~\ref{sec:kkt} with the duality-based bilinear formulation~\eqref{billinear_lower_l} and \eqref{billinear_lower_u} described in Section~\ref{subsec:duality} according to the
numbers of decision variables and constraints.

Both formulations involve bilinear terms but the bilinear formulation is more compact.
Consider a system with $b$ PQ buses, of which there are $r$ buses where the voltage magnitudes may violate their limits after the pre-screening described in~\ref{subsec:milp_reformulation}. The total number of  decision variables in the KKT formulation is $5\cdot b\cdot r + 2\cdot b$ ($2\cdot b\cdot r$ from power injections, $3\cdot b\cdot r$ from dual variables, $b$ from the voltage thresholds, and $b$ from the sensor locations). Our proposed duality-based bilinear formulation involves only $3\cdot b\cdot r + 2\cdot b$ decision variables. The reduction happens because the variables corresponding to power injections are entirely removed by duality.

Regarding the number of constraints, the duality-based bilinear formulation does not have the stationarity conditions, primal feasibility, or power injections directly involved, resulting in a reduction of $2\cdot b\cdot r + b\cdot r + 2\cdot b\cdot r = 5\cdot b \cdot r$ constraints. The implications of these differences on tractability is assessed via the solution times presented in Section~\ref{sec:simulation}.

\end{document}